# Atom probe characterisation of segregation driven Cu and Mn-Ni-Si co-precipitation in neutron irradiated T91 tempered-martensitic steel

T. P. Davis[*,1], M. A. Auger[1,2], N. Almirall[3], P. Hosemann[4], G. R. Odette[3], P. A. J. Bagot[1], M.P. Moody[1] and D. E. J. Armstrong[1]

[1]Department of Materials, University of Oxford, Parks Road, Oxford, OX1 3PH, UK
[2]Department of Physics, Universidad Carlos III de Madrid, Leganés, Madrid, Spain
[3]Materials Department, University of California, Santa Barbara, CA 93106, USA
[4]Department of Nuclear Engineering, University of California, Berkeley, CA 94720, USA

*Corresponding Author: thomas.davis@materials.ox.ac.uk

## ABSTRACT

The T91 grade and similar 9Cr tempered-martensitic steels (also known as ferritic-martensitic) are leading candidate structural alloys for fast fission nuclear and fusion power reactors. At low temperatures (300 to 400 °C) neutron irradiation hardens and embrittles these steels, therefore it is important to investigate the origin of this mode of life limiting property degradation. T91 steel specimens were separately neutron irradiated to 2.14 dpa at 327 °C and 8.82 dpa at 377 °C in the Idaho National Laboratory Advanced Test Reactor. Atom probe tomography was used to investigate the segregation driven formation of Mn-Ni-Si-rich (MNSPs) and Cu-rich (CRP) co-precipitates. The precipitates increase in size and, slightly, in volume fraction at the higher irradiation temperature and dose, while their corresponding compositions were very similar, falling near the Si(Mn,Ni) phase field in the Mn-Ni-Si projection of the Fe-based quaternary phase diagram. While the structure of the precipitates has not been characterized, this composition range is distinctly different than that of the typically cited G-phase. The precipitates are composed of CRP with MNSP appendages. Such features are often observed in neutron irradiated reactor pressure vessel (RPV) steels. However, the Si, Ni, Mn, P and Cu solutes concentrations are lower in the T91 than in typical RPV steels. Thus, in T91 precipitation primarily takes place in solute segregated regions of line and loop dislocations. These results are consistent with the model for radiation induced segregation driven precipitation of MNSPs proposed by Ke et al. Cr-rich alpha prime (α') phase formation was not observed.

## 1. INTRODUCTION

The ASTM Grade 91 tempered-martensitic (also known as ferritic-martensitic) 9-Cr and 1-Mo (91) steels are commonly known by their form names: T91[1], F91[2], P91 [3], where T, F, and P indicate tube, forging and pipe, [4]. Other names, which have been used for these steels, include Mod 9Cr-1Mo (United States) and 1.4903 X10CrMoVNb9-1 (Europe). More generally, we will refer to these alloys as 9 Cr normalized and tempered martensitic steels (TMS). Grade 91 (referred hereafter as T91) steel is a candidate for sodium [5] and lead/lead-bismuth [6] cooled advanced nuclear reactors. Similar reduced activation variants are candidates for structural components in future fusion power reactors [7]. The attractive properties of T91 steels for these applications include: a) excellent void swelling resistance [8–10]; b) high thermal conductivity and low thermal expansion; c) existing supply chain for steels widely used in boiler tubes, heat exchangers and piping [11] and; d) American Society of Mechanical Engineers (ASME) BPVC Section III Division 5 [12] nuclear code qualification.

Neutron irradiation drives microstructural and microchemical evolutions in TMS, like T91, which have detrimental effects on the mechanical properties, thus limiting the lifetime and performance of reactor structural components. Reported microstructural features that result from neutron irradiation of 9-12 wt% Cr tempered-martensitic alloys, such as T91, HT9 and Eurofer97, include the Mn Ni Si precipitates (MNSPs) [13–20] (often described as 'G-phase' – $Mn_6Ni_{16}Si_7$ [21]), Cr-rich alpha prime (α') precipitates [14,17,22–24], voids, dislocation loops, evolved network dislocations, and solute segregation to, and precipitation at, dislocations [19,25–27]. These microstructural changes are due to the excess radiation defect generation and clustering, dislocation climb, radiation-induced segregation (RIS) and radiation enhanced diffusion (RED). These hardening features embrittle the steels, as manifested by elevations of ductile-to-brittle transition temperature, increases in yield stress, decreases in ductility and degradation of fracture toughness [5].

A number of characterization studies of the MNSPs that form during neutron irradiation of T91 steel have been reported previously [13,14,18,28–33]. MNSPs also have been observed in ion [19,30,33–35] and proton [13,36–39] irradiated T91 and HT9 steels also. In all cases, MNSPs were associated with segregation of Mn, Ni, and Si at dislocations and Cu precipitates between temperatures of ~ 270 °C to ~450 °C. Jiao et al. [40] investigated MNSPs in a T91 steel neutron irradiated to 17.1 dpa and 35.1 dpa at several temperatures between 376 °C to 524 °C in the Russian BOR60 sodium-cooled fast test reactor. Segregation and MNSP precipitation were observed in all cases except at the highest temperature of 524 °C. Impurity Cu precipitates, primarily on dislocations enhance MNSP formation in 9Cr TMS, likely be promoting nucleation [29,33,38]. Notably, the typical Mn-Ni-Si contents of these alloys are relatively low, lying either near, or below, the thermodynamic solvus boundary. Recently, a cluster dynamics model showed that solute segregation is required for heterogeneous nucleation and growth of MNSPs on dislocations in T91 steel [29].

In contrast to Jiao's results at 524 °C, Adisa et al. [33] recently reported very large MNSP mole fraction ($f$) of ≈ 1.4% in T91 that was neutron irradiated at 500°C to 3 dpa. CRPs and solute segregation were also observed. However, in this case the MNSPs include Fe and Cr contents nominally found by CAMECA Integrated Visualization and Analysis Software (IVAS)®, the atom probe tomography (APT) reconstruction software, were at a very high concentration of ≈ 89%. We believe that the Fe and Cr atoms associated with the reported $f$ are APT artefacts, and is discussed in an experimental method section of this study; further, both experimental and first principles assessment of the relevant phases indicated that the relevant MNSP phases did not contain significant amounts of these matrix solvent elements. The Adisa et al. paper also reports application of a cluster dynamics (CD) simulation, developed by H. Ke at al. [41] which was later used by J. Ke et al. (2018) [29] to model precipitation in T91 steel. Adisa et al. reported that the CD model predicts similarly (quite large) $f$ values for G-phase precipitates, based on the trace bulk T91 solute content of Ni, Mn and Si, not accounting for segregation. However, a thermodynamic analysis provided in this study shows that, in this case at low bulk solute concentrations, the system is highly undersaturated; further the predicted Ni in the G-phase MNSPs greatly exceeds the total available in the alloy. Thus, while the results by Adisa et al. [33] presented solute precipitation of trace amounts at 500°C is a useful, our paper clarifies that this observation is also highly driven by solute segregation at dislocations, as modeled by J. Ke et al. [29].

In addition to MNSP observation in Fe-Cr based alloys, neutron irradiation embrittlement of low alloy reactor pressure vessel (RPV) steels has been the subject of extensive basic research for 40 years [42,43]. Embrittlement is caused by irradiation hardening primarily due to precipitation of supersaturated Cu impurities [44–46] and/or Mn Ni Si solute atoms [47–50]. Supersaturated Cu can form Cu-rich precipitates (CRPs) under very long-term thermal aging near RPV service temperatures [51]; however, Cu precipitation kinetics are highly accelerated by neutron irradiation due to radiation enhanced diffusion (RED) [43,47,52]. The formation of so called late blooming MnNiSi precipitates (MNSPs) was first predicted by Odette in 1996 [42]. In Cu bearing steels, MNSPs form as appendages to CRPs and slowly grow to large fractions of the steel Mn+Ni+Si alloying element contents, which are typically 2 to 3 at.% [43]. MNSPs also develop in low Cu steels, although even trace amounts of Cu are known to be potent catalysts for their formation [53]. The first experimental proof of MNSP formation in irradiated RPV steels was reported in 2004 [54]; and, since then, MNSPs have been widely observed [16,47,50,51,55–58]. Notably, CRPs and MNSPs are well predicted by thermodynamic and kinetic modelling [45,46,59]. Kinetic lattice Monte Carlo models recently showed that the co-precipitated morphology observed in APT reconstructions are the result of an interplay between interfacial energies, diffusion paths, such as through the Cu cluster, and ordering energies [46]. Both experiments and physical models show that MNSPs will dominate RPV integrity issues for life extension of light-water reactors [50,53,55]. RPV steel studies have also revealed significant solute segregation to loop and network dislocations. The segregated dislocations are a favoured nucleation sites for heterogeneous nucleation of MNSPs [43,47,50,57,58], as are cascade generated solute cluster complexes. However, apparently random homogeneous nucleation is also frequently observed in RPV steels and is predicted by models at sufficiently high solute contents (supersaturations), particularly Ni.

Thus, the objective of this study is to build on the understanding of precipitation in RPV steels and the corresponding much more limited database for Fe-Cr alloy systems, including model binary alloys and TMS 9-12%Cr steels like T91 and HT9 [5,9,19,60]. The major differences between these two alloy systems are that the solute contents of TMS are typically much lower than in RPV steels (expect Cr), while the dpa doses are much larger and the temperatures somewhat higher. The primary significance of these differences is that solute segregation to and heterogeneous nucleation on dislocations is critically important, due to the low solute content in TMS alloys. Here we focus on APT characterization of TMS T91, irradiated to 2.14 and 8.82 dpa at 327 and 377 °C, respectively. Note, such lower temperature investigations of T91 steel have practical importance because the inlet temperature of a potential liquid metal cooled reactor could be as low as 320–375 °C (the specific temperature range is reactor design dependent) [5], thus exposing the lower cladding tubes wrapper and structural components to service conditions, associated with maximum TMS irradiation embrittlement [7].

## 2. EXPERIMENTAL METHODS

### 2.1. AS-RECEIVED MATERIAL

The chemical composition of the T91 steel heat used in this study, which is given in Table 1, meets the required standard. The as-received APT compositions (at.%) are also given in Table 1 (see section 2.3 for the experimental procedure). The C is lower in the APT data, since it primarily resides in unprobed coarse carbides (expect in a later carbide dataset). The other elements are generally similar with the exception of Ni, which is significantly higher in the APT data (due to significant Ni segregation to nanosized features, as later discussed). APT also detects trace amounts of dissolved Cu.

Table 1: T91 Bulk Chemical and APT Composition Measurement (averaged across multiple datasets).

| Element | Bulk (wt%) | Bulk (at%) | APT (at%) |
|---|---|---|---|
| C | 0.07 | 0.32 | 0.02 ± 0.01 |
| Mn | 0.47 | 0.47 | 0.41 ± 0.02 |
| P | 0.02 | 0.04 | 0.02 ± 0.01 |
| S | 0.02 | 0.04 | - |
| Si | 0.28 | 0.55 | 0.54 ± 0.01 |
| Cr | 9.24 | 9.84 | 8.81 ± 0.17 |
| Mo | 0.96 | 0.55 | 0.41 ± 0.09 |
| Ni | 0.16 | 0.15 | 0.39 ± 0.02 |
| V | 0.21 | 0.21 | 0.10 ± 0.01 |
| Al | - | - | 0.04 ± 0.01 |
| Cu | - | - | 0.03 ± 0.01 |
| Co | - | - | 0.01 ± 0.01 |
| Fe | Bal. | Bal. | 89.25 ± 0.34 |

## 2.2. NEUTRON IRRADIATION

The neutron irradiated alloys were irradiated in the Advanced Test Reactor (ATR): a) as part of the University of California Santa Barbara (UCSB) ATR-1 experiment and are included in the Nuclear Science User Facilities (NSUF) Library irradiation [61]; and, b) as part of the University of Illinois Urbana Champagne (UIUC)) experiment [62]. Both of these irradiations were drop-in experiments, which did not include thermocouples to directly monitor temperatures. Rather, the temperatures were regulated by a combination of insulating helium/argon mixture gas gap and nuclear heating The temperatures were estimated based on Abaqus thermal heat transfer and MCNP code [63] for nuclear heating analysis and the reactor lobe power history [61]. Specimen 0020-2008-139 from the UCSB-1 library and specimen 2008-92-387 from the UIUC library, both have similar compositions and are believed to be from the same heat provided by Los Alamos National Laboratory (see composition in Table 1).

Table 2: Irradiation Conditions for the T91 specimens in the ATR reactor [61].

| Specimen ID | Steel | Temp. (°C) | Neutron flux (n/cm$^2$/s, E > 1Mev) | Neutron fluence (n/cm$^2$ E > 1 MeV) | Dose (dpa) |
|---|---|---|---|---|---|
| UCSB 0020-2008-139 | T91 | 327 | $1.21\times10^{14}$ | $1.57\times10^{21}$ | 2.14 |
| UIUC 2008-92-387 | T91 | 377 | $2.30\times10^{14}$ | $\sim7.80\times10^{21}$ | 8.81 |

## 2.3. ATOM PROBE TOMOGRAPHY

As-received T91 steel (the same steel batch as 0020-2008-139 and 2008-92-387 sample ID) was analysed using the APT technique [64]. APT analysis on the as-received T91 steel was conducted with a CAMECA LEAP® 5000XR at the Department of Materials, University of Oxford. APT specimens were prepared by the lift-out technique [65] using a Zeiss Crossbeam 540 Analytical Focused-Ion Beam (FIB)-Scanning Electron Microscope (SEM).

Polished and mounted T91 TEM discs were provided by UCSB via the Nuclear Science User Facility (NSUF) of the US Department of Energy. The post neutron irradiation examination was conducted at the Microscopy and Characterization Suite located at the Center for Advanced Energy Studies (CAES) with the support from the NSUF. APT analysis was conducted with a CAMECA LEAP® 4000X HR. The APT specimens were prepared by the lift-out technique using a FEI Quanta 3D FEG FIB Scanning Electron Microscope (SEM). A final FIB cleaning process was performed by using 2 kV Ga ions, in order to minimise FIB-induced damage. The final milled specimen diameters were between ~50 - 100 nm.

In both cases, the APT specimens were analysed at a stage temperature of 55 K, a voltage pulse fraction of 20%, a pulse rate of 200 kHz and the average detection rate was set to 1.0 %. The detection efficiency of the LEAP® 4000X HR and LEAP® 5000 XR were 37% and 52%, respectively. CAMECA IVAS® version 3.8.4 was used to reconstruct all atom probe datasets. Calibration of the final reconstructed APT maps used SEM micrographs of the final tip shape and crystallographic pole indexing.

The search for MNSP was conducted by using both the maximum separation method [66] and the core-link method [67] with the following parameters (averaged): $D_{MAX}$ = 0.85 nm, Order = 2, $N_{MIN}$ = 35 and $D_{erosion}$ = 0.425 nm. These parameters were optimised following the method outlined by Williams, C.A. et al [68]. After the cluster search in IVAS was complete for all datasets (8 datasets with >5M ions) for 2.14 dpa and (7 datasets with >5M ions for 8.82 dpa), the sizes, number densities, compositions, and volume fractions of the precipitates were calculated, where the latter is based on the fraction of solute ions in the clusters. Partial edge clusters (defined as a cluster from the original material whose true extent is not completely contained within the reconstructed APT volume) were removed from the cluster search to avoid underestimation of the sizes. The algorithm used to detect and remove the edge clusters was developed by Jenkins et al. (2020) [58,69]. The number density of clusters, $N_d$, was calculated by the following:

$$N_d = \frac{N_{ClustersDetected} - \frac{1}{2} N_{EdgeClusters}}{V_{Dataset}} \quad (1)$$

where $N_{ClustersDetected}$, is the number of clusters within the analysed reconstruction and $N_{EdgeClusters}$ is the number of clusters at the edge of the reconstruction, and $V_{Dataset}$ is the volume of the reconstruction dataset (in $m^3$). The volume was determined by:

$$V_{Dataset} = \frac{N_{Ranged} \Omega}{\eta} \quad (2)$$

Here $\Omega$ is taken as the volume of one Fe atom ($1.178 \times 10^{-2}$ nm$^3$), $N_{Ranged}$ is the ranged atoms within the APT dataset, and $\eta$ is the detection efficiency of the atom probe instrument used. The volume fraction, f, of the MNSP were calculated by:

$$f = \frac{N_{Ranged}^{Cluster} - N_{Fe}^{Ranged}}{N_{Total}}, \quad (3)$$

where $N_{Ranged}^{Cluster}$ is the number of ranged atoms within all clusters, $N_{Fe}^{Ranged}$ is the number of ranged Fe in all clusters and $N_{Total}$ is the total number of ranged atoms within the dataset. The volume of each cluster was assumed to be spherical and with the atomic density of bcc-Fe.

The atom probe mass-to-charge-state spectrum must be ranged by user assigned chemical identities to each peak. With steels that have Ni and Si alloying addition, overlapping mass-to-charge-state peaks occur at 29 Da with $^{58}Fe^{2+}$, $^{58}Ni^{2+}$ and $^{29}Si^{1+}$ with all three ion species potentially incorporated into the MNSPs. The contribution of $^{58}Fe^{2+}$ to the 29 Da peak with the clusters has previously been claimed to be limited in other irradiated steels [70]. Some authors have chosen not to include any ions originating from the 29 Da peak in their solute cluster definitions [49]. Other researchers have utilised Scanning TEM (STEM)-Energy-dispersive X-ray spectroscopy (EDS) to claim that the exclusion of Fe from the defined MNSP precipitates in the APT data was reasonable [71]. The situation in APT is further complicated by trajectory aberrations that affect the spatial resolution of the reconstructed atom maps, as discussed by Larson et al. [72], which can erroneously introduce Fe from the surrounding matrix into the defined MNSP cluster. Therefore, in this study, all cluster compositions did include the peak at 29, identifying this as $^{58}Ni^{2+}$ (not $^{29}Si^{1+}$ as it did not match the expected natural isotope ratios) and the Fe ions will be removed from the cluster calculation, which is similar to previous studies [53,58,71]. For an additional check, the spatial distribution of ions within the reconstructed data associated to the 29 Da was visually inspected and observed to correspond to the locations of MNSPs, further indicating that the majority of 29 Da peak was attributed to $^{58}Ni^{2+}$.

Following the procedure by N. Almirall et al. [73] solute segregation to dislocations, 2.0 at% Si isoconcentration surface was created and a region of interest (ROI) was placed through the dislocation's core in both transverse and longitude direction to generate composition line profiles.

## 3. RESULTS

### 3.1. AS-RECEIVED T91 STEEL

A typical reconstruction of the as-received T91 steel is shown in Figure 1(a). The microstructure was homogeneous and presented no evidence of nanometric sized ppts or solute segregation.

### 3.2. NEUTRON IRRADIATED T91

The typical APT reconstruction in Figure 1(b) of the neutron irradiation T91 steel at 2.14 dpa at 327 ºC, shows the formation of Cu rich clusters (CRPs) as well as, segregation of Si, P and Ni to dislocations. The corresponding MNSPs on the dislocations were found to be appendages to CRPs (shown later in Figure 4 and Figure 6). No clustering of Mo, V, Co, Al, Fe or Cr was detected. The average composition of MNSPs was calculated for each dataset and is shown on a ternary projection of the Fe-Mn-Ni-Si phase diagram in Figure 2. The average of 8 APT datasets of MNSP-CRP volume fraction, average radius, volume, composition and number density are summarized in Table 3. Region of Interest (ROI) solute segregation profiles, both transverse and longitudinal (along) dislocations, are shown in Figure 4. The longitude solute profile in Figure 4(c), clearly shows the periodic formation of MNSPs-CRPs along the dislocation line. Enrichment of Ni, Si and P at a carbide interface is shown in Figure 5.

A typical APT reconstruction of T91 neutron irradiated to 8.8 dpa is shown in Figure 1(c). As seen in the close-up atom map, the MNSPs are appendages to CRPs, as frequently observed in RPV steels [53,59] and proton irradiated T91 [34]. No clustering of Mo, V, Co, Al, Fe or Cr was detected. The average compositions of the MNSPs were calculated for each tip dataset and are shown on a ternary phase diagram projection in Figure 2. The average MNSP volume fraction, radius, volume, composition and number density measured for all 7 APT datasets for this irradiation condition of MNSP volume fraction, average radius, volume, composition and number density are summarized in Table 3. The APT reconstruction in Figure 1(c) suggests that the MNSP-CRP features have slightly larger and better defined volumes in the T91 steel neutron irradiated to 8.82 dpa at 377 °C than in the 2.14 dpa at 327 °C condition, along with less apparent segregation of Si, P and Ni to dislocations. Solute segregation to dislocation loops in Figure 6 is shown as transverse concentration profiles. The profile in Figure 6(b) is a 1.0 at.% Cu isoconcentration surface marking a CRP. The longitudinal solute segregation profile, shown in Figure 6(d), indicates the formation of MNSPs-CRPs at the edge of a dislocation loop.

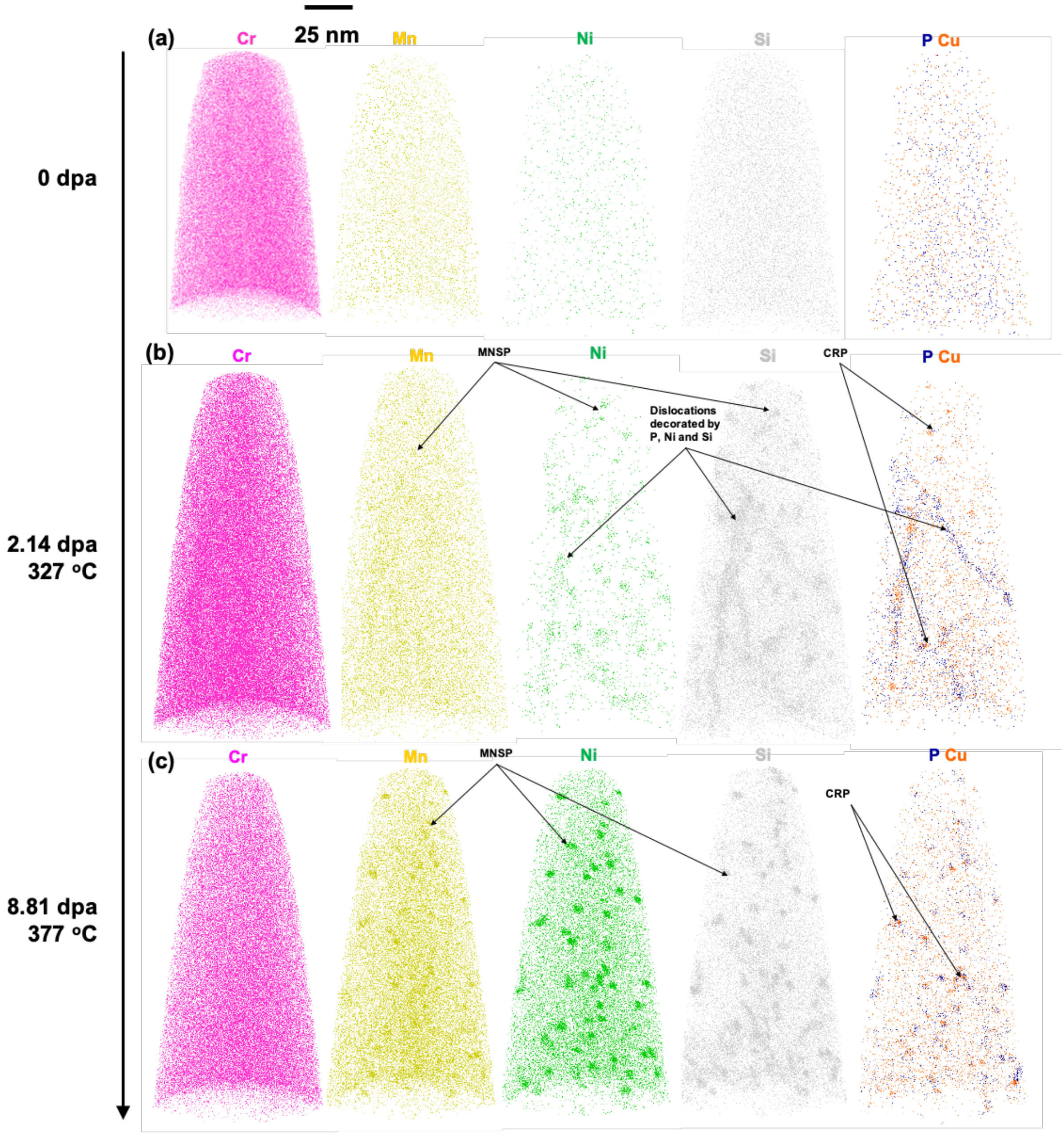

Figure 1: (a) APT reconstruction of the as-received T91 steel, showing a homogeneous microstructure; (b) APT reconstruction of the T91 steel neutron irradiated to 2.14 dpa at 327 °C showing MNSPs and CRPs, as well as Ni/Si/P segregating to dislocations; (c) APT reconstruction of the T91 steel neutron irradiated to 8.82 dpa at 377 °C, showing better defined MNSPs and P/Cu precipitates, perhaps with somewhat less Ni/Si/P segregation to dislocations (but visible in Si). Cr appeared homogeneously distributed in all conditions.

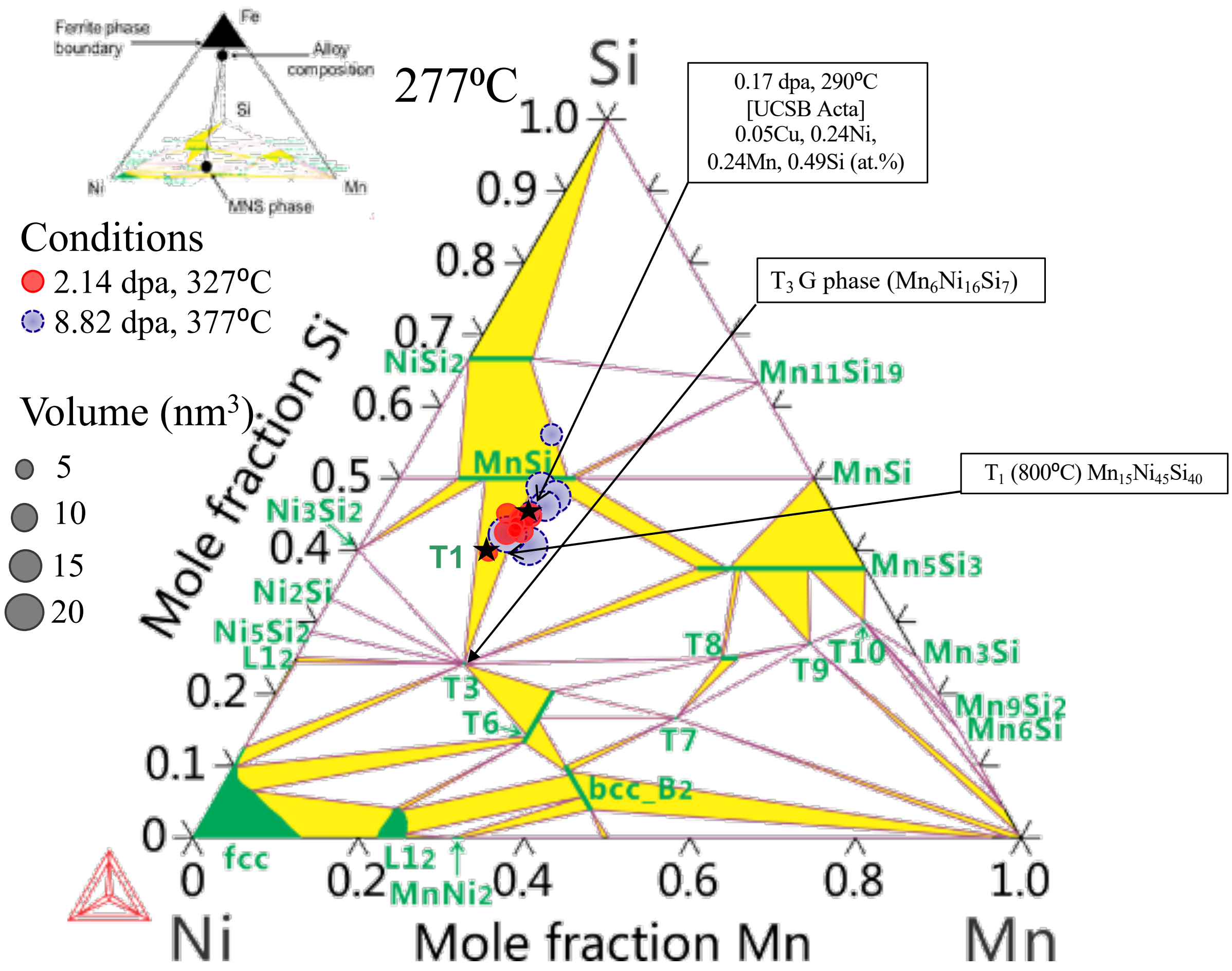

Figure 2: The averaged MNSP composition (in at%) of each APT dataset (over >5 million ions) for both 2.14 dpa and 8.82 dpa T91 irradiation conditions displayed on a 277 °C isothermal section of the Mn-Ni-Si ternary system projection of a Fe based phase diagram [41,48,74]. The size of the data point is scaled to the MNSP average APT dataset volume (nm$^3$). Phase $T_1$ (800°C) $Mn_{15}Ni_{45}Si_{40}$ from [75] for comparison. The stoichiometric 'G-phase' phase, $Mn_6Ni_{16}Si_7$, is marked as T3. Low Ni containing RPV steel neutron irradiated to 0.17 dpa at 290 °C is provided from [53] for comparison.

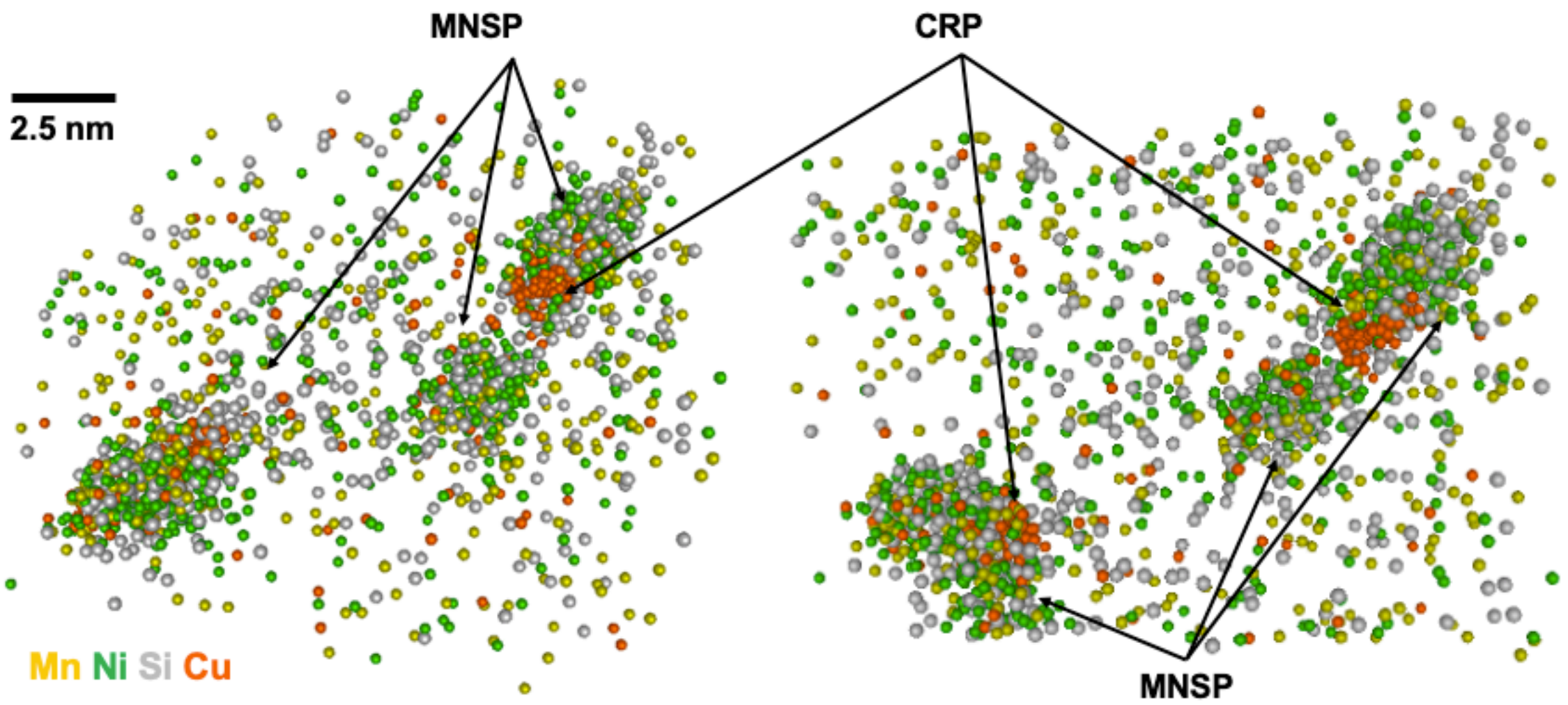

Figure 3: An APT reconstruction closeup view of MNSPs and CRPs in T91 neutron irradiated to 2.14 dpa at 327 °C as shown in Figure 1 (b). The right-hand reconstruction is a 180° rotation of the left-hand reconstruction to provide both views.

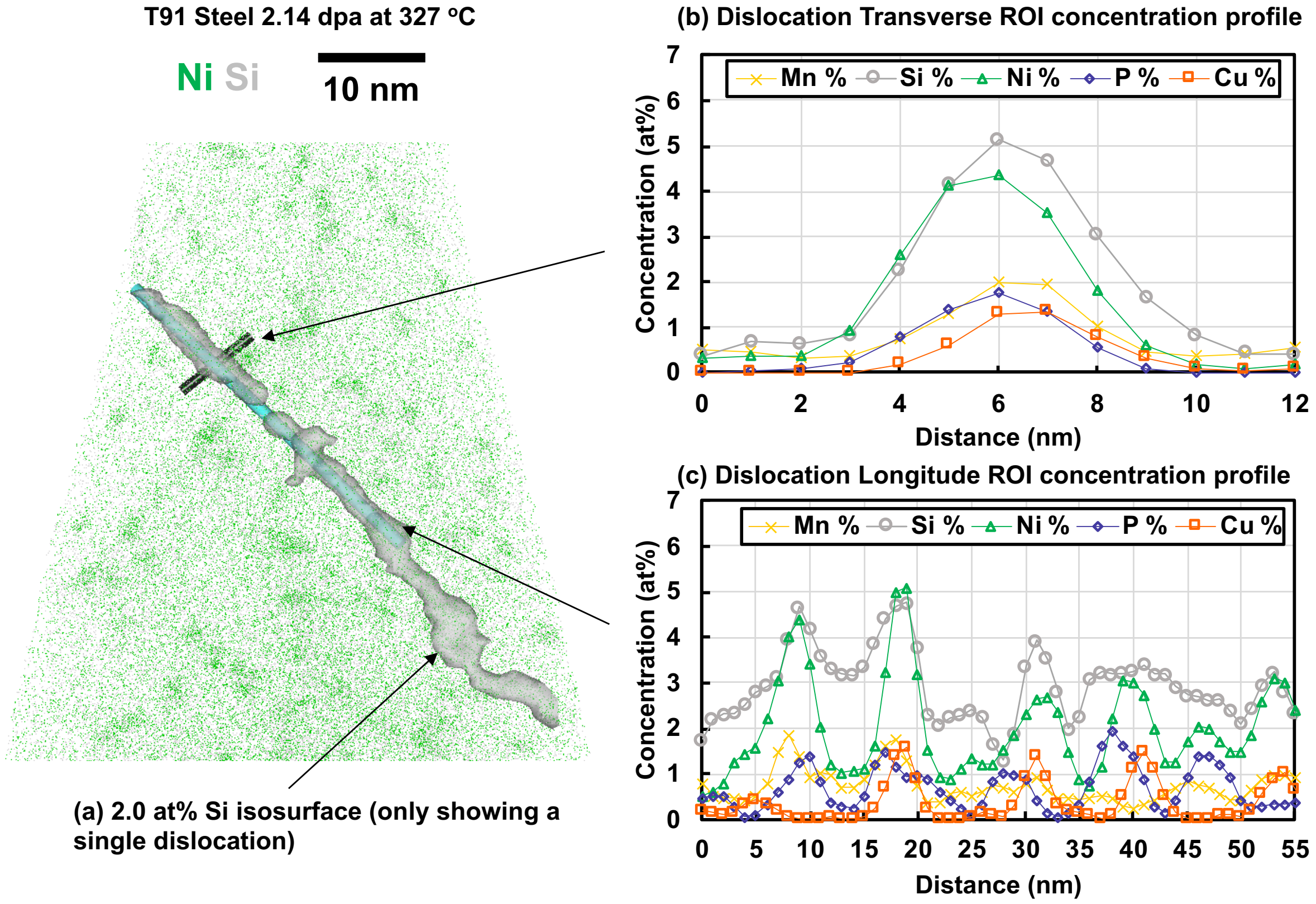

Figure 4: Solute segregation to a line dislocation in a T91 steel irradiated to 2.14 dpa at 327 °C. (a) displays a 2.0 at% Si isoconcentration surface; (b) is a 1D transverse ROI concentration profile at a random section of the dislocation and (c) is a 1D longitude ROI concentration profile inside the dislocation displaying peaks in Si, Ni, Mn, P, and Cu precipitation (and Cu is shifted to the right indicating MNSP appendage to CRPs).

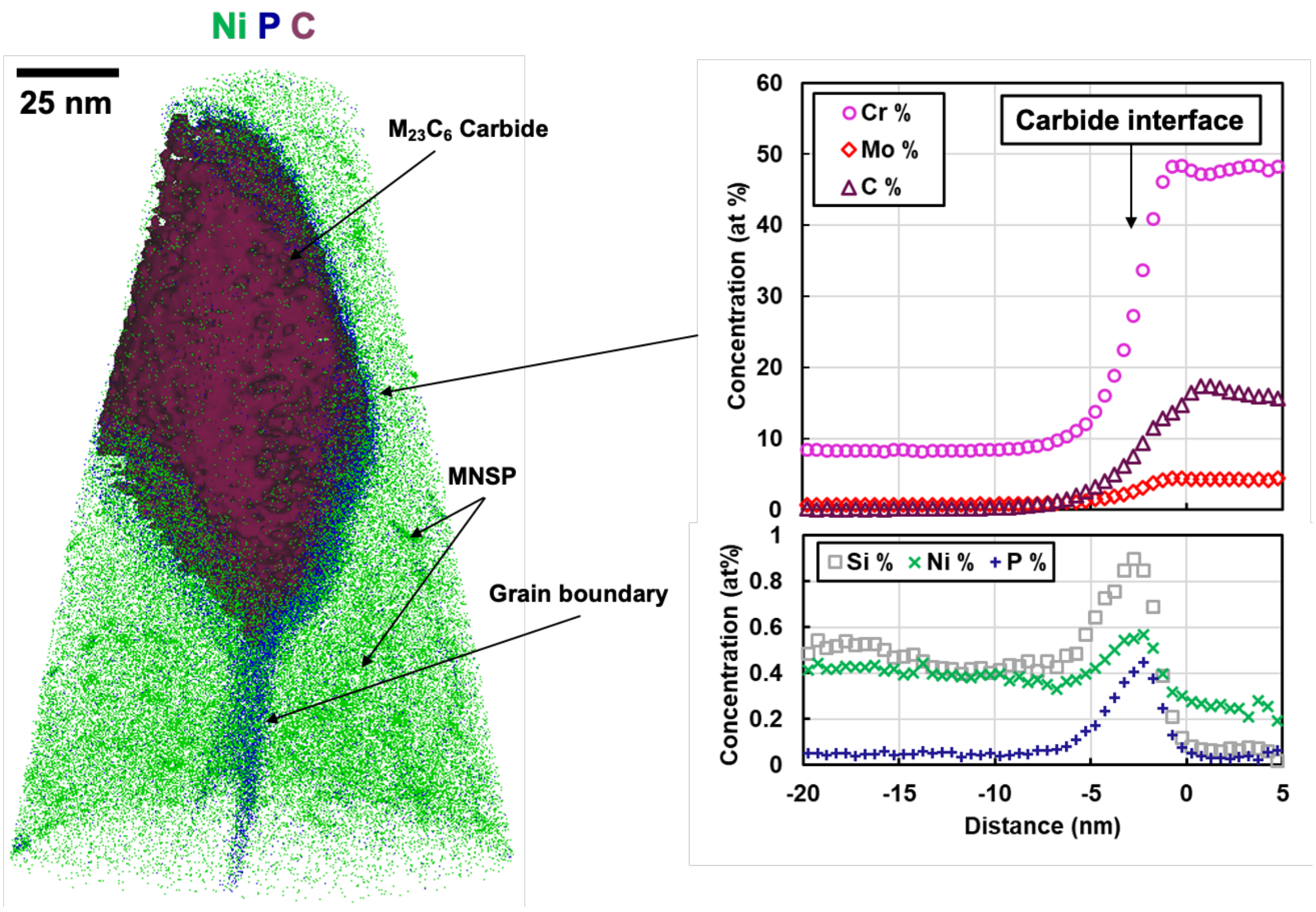

Figure 5: An APT reconstruction of T91 neutron irradiated to 2.14 dpa at 327 °C. A chromium-based carbide can be seen with a grain boundary/interface that is decorated with Ni, Si and P. The concentration profile was produced using a proxigram from a 5 at% C isoconcentration surface.

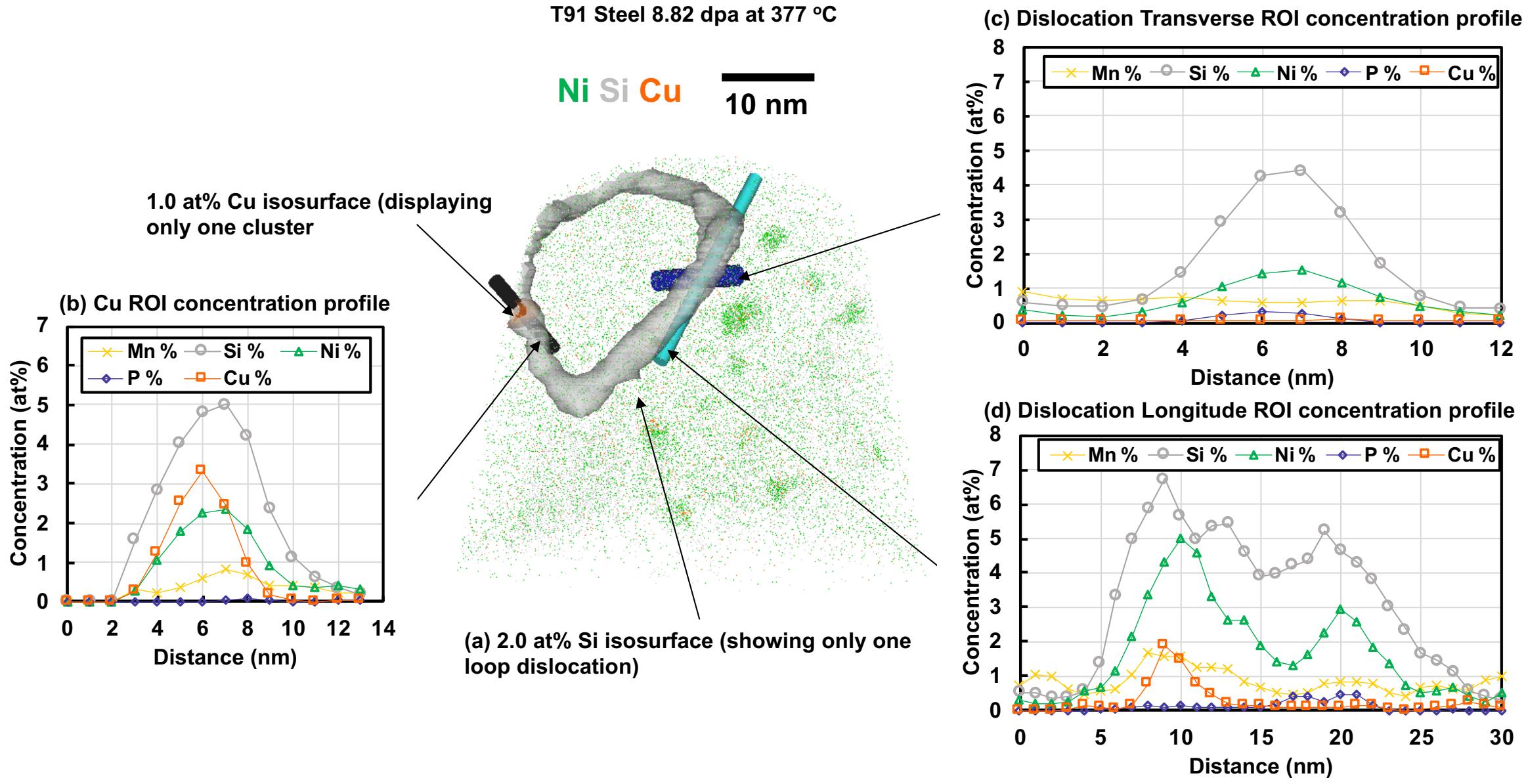

Figure 6: Solute segregation to a dislocation loop in a T91 steel irradiated to 8.82 dpa at 377 °C; (a) displays a 2.0 at% Si loop isoconcentration surface; (b) is a 1D concentration profile through the 1.0 at% Cu isoconcentration surface; (c) is a 1D transverse ROI concentration profile centered on a random section of the dislocation loop and (d) is a 1D longitude ROI concentration profile inside the randomly selected segment of the dislocation loop.

Table 3: The volume fraction, average diameter, average volume, volume fraction and solute composition of MNSP in both neutron irradiation conditions using the cluster search, as outlined in Section 2.3.

| | Irradiation Conditions | |
|---|---|---|
| | 2.14 dpa at 327 °C | 8.82 dpa at 377 °C |
| **MNSP (only solute ions, excluding Fe)** | | |
| Number Density (#/cm$^3$) | (3.1 ± 0.7) $10^{23}$ | (1.5 ± 0.7) $10^{23}$ |
| Average Radii (nm) | 1.1± 0.1 | 1.45 ± 0.2 |
| Average Volume (nm$^3$) | 5.8 ± 1.3 | 15.6 ± 5.4 |
| Volume fraction (%) | 0.26 ± 0.06 | 0.33 ± 0.14 |
| **MNSP solute compositions (Fig. 6)** | | |
| Mn (%) | 16.49 ± 4.7 | 16.47 ± 4.4 |
| Ni (%) | 37.87 ± 6.33 | 36.80 ± 10.61 |
| Si (%) | 45.64 ± 7.62 | 46.73 ± 11.59 |

## 4. DISCUSSION

The defining microstructural features of T91 steel are the martensitic lath microstructure, high network dislocation densities, $M_{23}C_6$ carbides, as well as somewhat finer scale vanadium nitrides and molybdenum/niobium carbides [60]. The latter features provide the high creep strength [76] needed for elevated temperature boiler and nuclear applications [9]. Note these performance-enabling microstructures are unstable in T91 under irradiation at temperatures more than 450 to 500°C, as indicated by irradiation softening and non-hardening embrittlement [77]. At lower temperatures, the main effects of irradiation are hardening and embrittlement due to segregation and precipitation of solutes, as described in the previous Section. It is well known, and the topic of a large amount of literature, that a smaller amount of hardening is contributed by dislocation loops, as illustrated in [24] for Fe-Cr binary alloys. Potential contributions from the evolution of network dislocations is rapidly emerging as an issue, but is not close to being well quantified [26,27,40,43,78]. However, discussion evolved dislocation loop and line hardening, including the effects of segregation and precipitation, is beyond the scope of this paper. Here we focus on high mass and spatial resolution APT to investigate segregation and nanometric precipitates, which are absent prior to neutron irradiation.

Precipitate evolution under neutron irradiation was illustrated in Figure 1 (a) – (c). Cr remains homogeneously distributed in all conditions. The high number density nanosized MNSPs act as dispersed barrier obstacles dislocation glide, which results in the hardening and shifts in the ductile-to-brittle transition temperature [5,9,60,79]. The decrease in number density and increase in size (and individual precipitate volume) in the higher dpa and temperature irradiation condition is expected based on the thermo-kinetics of segregation and precipitation; the slightly larger volume fraction at higher fluence and temperature (from 0.26 ± 0.06 to 0.33 ± 0.1 %) is probably within the uncertainties in the APT measurements. Note, RPV studies show that dispersed barrier hardening and embrittlement are primarily controlled by the square root of the precipitate volume fraction [50,53,55,57]. The T91 volume fractions correspond to estimated hardening contributions of ≈168 to 186 MPa.

The average composition of MNSP in both irradiated conditions, shown in Figure 2 and Table 3, is significantly different that the most often cited 'G-phase'. Rather, the MNSP compositions are closer to the Si(Mn,Ni) phase field in the MnNiSi ternary projection of the Fe matrix-based quaternary phase diagram. This phase field, calculated by Xiong et al., is $Si_{0.5}Mn_xNi_{(1-x)}$ where x varies from ~0.1 to 0.25 [48]. Note, the crystallographic structure of the precipitates in this study has not been characterized. Similar Si(Mn,Ni) phase compositions have been observed on dislocations in a low Ni (0.07 at.%) VVER-440 RPV steel [80]; and more recently in another low Ni RPV steel [57].

The T91 2.14 dpa at 327 °C carbide shown in Figure 5 is likely a $M_{23}C_6$ phase with a composition of ~ 48.2 ± 0.5 Cr, 16.7 ± 0.4 C, 4.3 ± 0.3 Mo and 29.0 ± 0.5 Fe (in at.%); however the crystallography has not been determined. The composition line profile across the carbide/matrix interface in Figure 5 shows enrichment of Ni, Si and P, which correlates well with previously analysed irradiated carbide interfaces [31,32,37].

Co-precipitation of CRPs and MNSPs, clearly seen in Figure 3, is widely observed in RPV steels at high fluence, and has been extensively characterized [47–50] and modelled [43,45,46,59]. CRPs have been observed in neutron and ion irradiated T91 [39,40] and HT9 [19,31]. Indeed, Cu driven co-precipitation has also been exploited by high strength steels to promote the formation of various intermetallic phases [81]. The corresponding sequence-of-events begins with the rapid precipitation of highly supersaturated Cu, with shells composed of the other solutes. After Cu is depleted from the matrix, Mn, Ni and Si continue to accumulate, so as to eventually form a separate appendage phase [42,43,46,53,59,82]. Even trace amounts of Cu act as a powerful catalyst for MNSP formation [53]. In the case of T91 co-precipitation takes place in highly solute segregated regions at dislocations, as previously observed by APT in irradiated T91 steel [19,39,40].

The nominal transverse solute concentrations profiles at line dislocations in Figure 4(a,b) reaches 4-5% for Si and Ni, and 1 to 2% for Mn, P and Cu. Figure 4(c) presents longitudinal profiles that show periodic peaks of Mn, Ni, Si, Cu and P, indicating the formation of precipitates. Note, the Cu peaks are slightly displaced from those for Ni, and especially Mn, which is consistent with co-precipitation [46]. The Si enrichment is very high and more uniformly distributed along the dislocation lines. This may rationalize the $SiMn_xNi_{(1-x)}$ precipitate compositions, which are near the Si(Mn,Ni) phase field, as shown in Figure 2. As seen in Figure 6, generally similar solute segregation also occurs at dislocation loops. However, P does not appear to segregate to loops. Cu is localized in the precipitate regions of both dislocation features. Note, these solute compositions are nominal, and may be affected by APT artefacts like trajectory aberrations.

Our analyses highlight the parallels that can be drawn between the much more limited database on MNSPs in neutron irradiated for Fe-Cr alloy systems, including model binary alloys and 9-12%Cr steels like T91 and HT9, with the much more extensive literature on precipitate evolution in RPV steels [16,41,52,55–57,59,42,43,46–51]. Notably, precipitates in some very low Ni RPV steels have compositions that fall near the Si(Mn,Ni) phase field. The major difference between these alloy classes is that the typical Cu + Ni + Mn +Si solute contents in RPV steels are much higher (≈ 3%) than in the 9Cr TMS alloys (≲1%). Thus, while precipitation at segregated dislocations (loops and line) occurs and is important for some RPV steels and irradiation conditions, local solute enrichment is probably necessary in alloys like T91. Note, segregation may be either thermally driven, or induced by irradiation (RIS), or both. The main thermodynamic driver for co-segregation is the local bonding interactions between the solutes, lowering their free energies near the dislocations. The other difference between RPV and 9Cr TMS, is that service conditions for the latter involves much higher temperatures and dose.

A combined solute segregation and cluster dynamics (CD) model has been used to predict the nucleation and growth of MNSPs in a subsaturated T91 steel as a function of irradiation dose in dpa [29]. The model, which predicts the MNSP number density, volume fraction and mean radius (and radius distribution), was previously calibrated using the results of a single proton irradiation at 400 °C to 7 dpa [37]. Figure 7 compares these predictions to the 8.8 dpa at 377°C APT data in this study, showing that the model predictions are in agreement with the neutron results. The number density and radius are in almost exact agreement, while the volume fraction is slightly under predicted, perhaps partly due to the lower neutron irradiation temperature. Figure 8 compares the predictions of solute segregation to the observed values; the agreement is reasonable in the case of Si, but Ni segregation is under predicted. It is likely that this is due to the fact that the model does not treat co-segregation of solutes. That is, solute-so lute interactions in the semi-dilute local micro-alloy regions at dislocations lower free energies in the segregated regions at dislocations. The Cu, Mn and P segregation that is observed in the T91 data is also shown but has not been modelled.

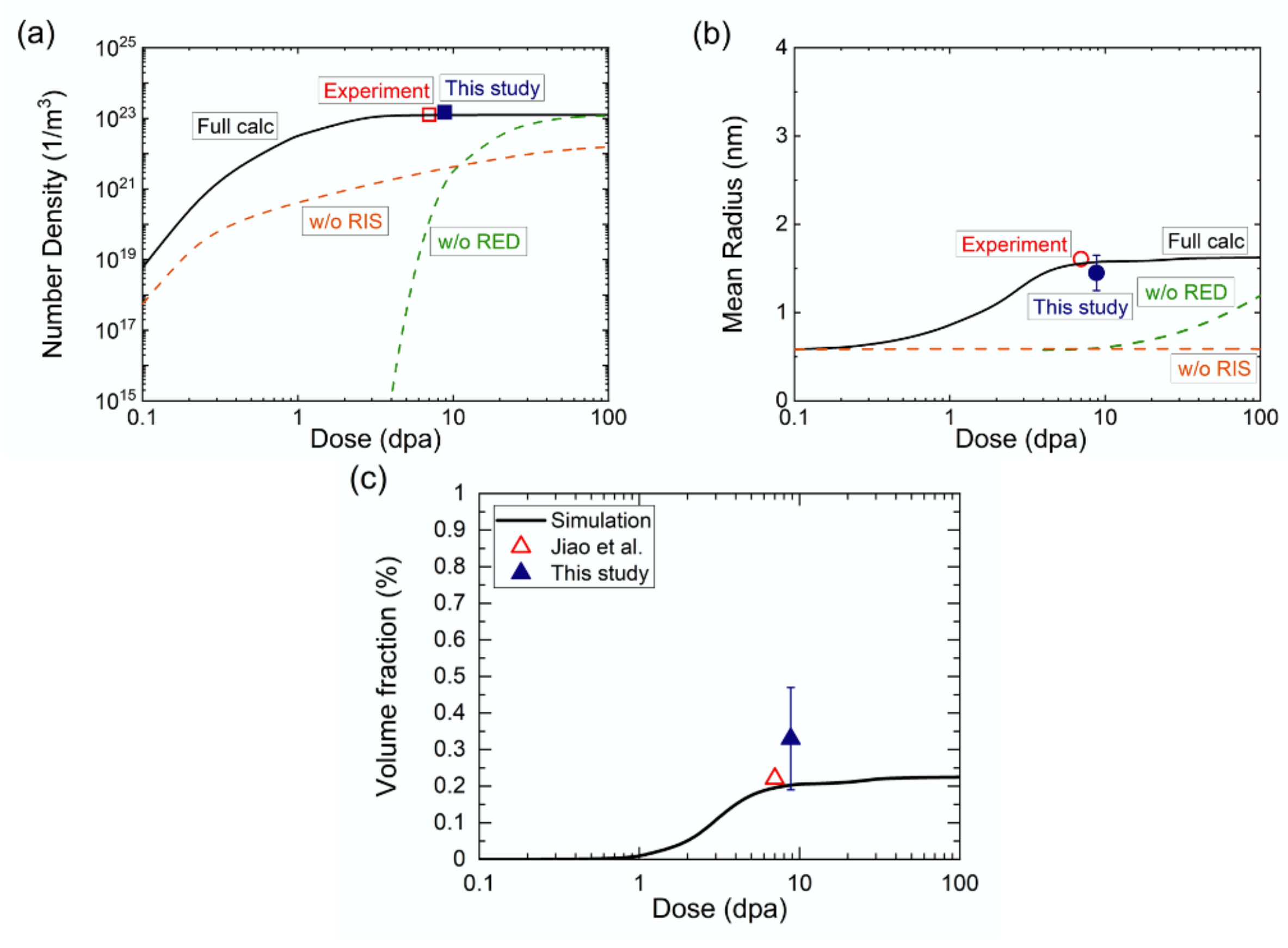

Figure 7: The comparison between the T91 MNSP (a) number density, (b) radius, and (c) volume fraction with the full calculation (segregation plus CD) model developed by J. H. Ke et al. [29]. The absence of precipitation without RIS segregation is also shown along with the sluggish kinetics without RED. Permission for reproducing J. H. Ke et al. [29] data has been granted.

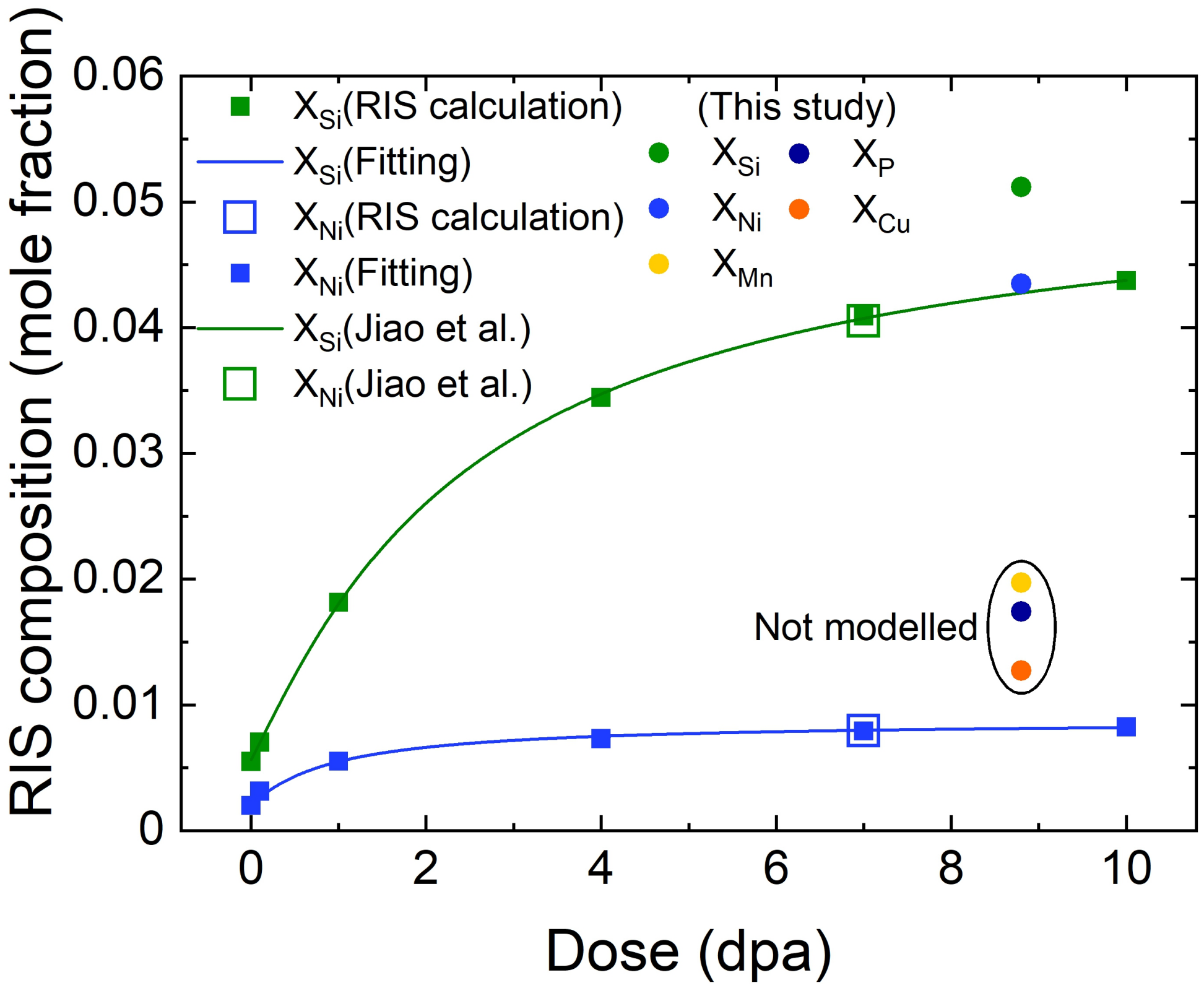

Figure 8: A comparison between T91 Si and Ni segregation to dislocations observed with predictions by J. H. Ke et al. [29]. The observed segregation of Cu, Mn and P is also show but was not modelled in the original study. Permission for reproducing the figure has been granted.

Notably, distinct α' precipitates were not observed in either irradiation condition. The Cr solubility limit is ≈ 8.8 ± 0.5 at% [22,83,84] at 300°C. Thermal α' precipitation can occur but is sluggish. However, Cr precipitation is greatly accelerated by radiation-enhanced diffusion [85]. The absence of α' in T91 is due likely to the low, or absent, Cr supersaturation, depending on the temperature. Such low supersaturations are insufficient to form significant populations of small, discrete α' precipitates. Results in the literature on α' vary [22,23,84,86]; for example, SANS [87] showed that a T91 alloy neutron irradiated to 0.7 dpa at 325 °C in the OSIRIS reactor contained a small volume fraction (≈ 0.1%) of high number density ($9.0 \times 10^{23}$ $m^{-3}$ ) α′ precipitates with an average radius of 1.3 nm [87]. Conversely, SANS and TEM did not observe α′ in a T91 steel irradiated to 184 dpa since it was irradiated at a higher temperature of 413 °C in the Fast Flux Test Reactor [14]. Further discussion of this widely studied topic is beyond the scope of this paper.

While not the primary focus of this work, it is useful to briefly discuss the thermodynamics that appear to be at work here. As noted previously, we have not measured the structure of the MNSPs. However, our results show that the compositions fall near the Si(Mn,Ni) phase field (as calculated by Xiong et al. [48]), and as seen in Figure 2. Further, Hu et al. [75] reported the existence of a $Mn_{15}Ni_{45}Si_{40}$ compound $T_1$ phase in the 800°C Mn-Ni-Si isotherm, also shown in Figure 2. These compositions bracket those found in this study, and at least one other on a similar T91 at 500°C and 3 dpa [33].

A full thermodynamic assessment of these phases is not yet available. However, the bulk and segregated alloy compositions can be compared to the thermodynamic solute product requirements for G-phase formation by the reaction the reaction:

$$16Ni_s + 6Mn_s + 7Si_s \rightleftharpoons Ni_{16}Mn_6Si_7 \qquad (1)$$

where the subscript s denotes that the solutes are in solution. Thus, the reaction solute product is $SP = \{[X_{Ni}]^{16}[X_{Mn}]^{6}[X_{Si}]^{7}\}^{1/29}$, where $X_i$ are the mole fractions of the dissolved Ni, Mn and Si. The equilibrium solute product ($SP_e$), or phase boundary, at 377°C is ≈ $5x10^{-3}$ [41]. The APT composition $X_i$ values in Table 1 yield an alloy solute product of ≈ $4.3x10^{-3}$; thus, the bulk system is nominally slightly undersaturated. In contrast, the local peak segregated $X_i$ for Mn ≈ 0.01, Si ≈ 0.05 and Ni ≈ 0.02 , in Figure 6(c), produces a local alloy SP of ≈ 0.044, which is highly supersaturated. The difference between the composition of G-phase, and that found in this work, is almost certainly due to the unusually high Si concentration in the segregated region. As shown by J. Ke et al. (2018) [29], RIS results in Ni and Si segregation, which is sufficient to drive precipitation. However, it should be noted that thermal segregation and precipitation, which is greatly accelerated by RED, could also be significant [88–90].

## 5. CONCLUSIONS

APT was used to quantify and understand the effect of neutron irradiation on solute segregation and CRP and MNSP co-precipitation in T91 steel. Two ATR irradiation conditions were investigated: 2.14 dpa at 327 ºC and 8.82 dpa at 377 ºC. Key observations are as follows:

- The compositions of all the MNSPs were similar and fell near the Si(Mn,Ni) phase field. While the structure of the precipitates has not been characterized, this composition range is distinctly different than for that for the typically cited G-phase and is consistent to the observations of Si(Mn,Ni) precipitates in neutron irradiated low Ni % RPV steels.
- Co-precipitation of MNSP and CRP is observed, where the MNSPs appear as an appendage to the CRPs.
- Significant solute segregation (P, Si, Ni and Mn) to dislocation lines and loops is observed, with large enrichment factors for Si and Ni ≈ 10.
- MNSP and CRP form on the microalloyed regions at dislocation lines and loops.
- The T91 bulk Ni, Mn and Si composition is undersaturated and insufficient to cause precipitation. However, the corresponding composition in microalloy regions of segregated dislocations is highly supersaturated.
- CRP-MNSP number densities, sizes, volume fractions, and Si enrichment at dislocations are in good agreement with predictions of a previously reported model, combining solute segregation and CD submodels. Ni solute segregation to dislocations is significant but under predicted in the J. Ke model.
- α' formation was not observed in any atom probe data sets.

This study has provided an insight into the MNSP compositions, volume fractions and sizes, which may contribute to a better understanding of the embrittlement of T91 steel. Moreover, this study builds upon the extensive understanding of precipitation in RPV steels and corresponding much more limited Fe-Cr alloy systems.

## ACKNOWLEDGEMENTS

T. P. Davis is funded by the Clarendon Scholarship from the University of Oxford and Engineering and Physical Sciences Research Council Fusion Centre for Doctorial Training [EP/L01663X/1]. APT was supported by EPSRC grant EP/M022803/1 “A LEAP 5000XR for the UK National Atom Probe Facility.” The authors acknowledge use of characterisation facilities within the David Cockayne Centre for Electron Microscopy, Department of Materials, University of Oxford, alongside financial support provided by the Henry Royce Institute (Grant ref EP/R010145/1). This work was supported by the U.S. Department of Energy, Office of Nuclear Energy under DOE Idaho Operations Office Contract DE-AC07- 051D14517 as part of a Nuclear Science User Facilities (NSUF) experiment. Post irradiation experiments on neutron irradiated samples were conducted at the Microscopy and Characterization Suite (MaCS), Center for Advanced Energy Studies using the LEAP 4000X HR under the NSUF Rapid Turnaround Experiment No 19-1721. The deep, long-term involvement of UCSB in this work was supported by the DOE National Scientific Users Facility, who sponsored the ATR-1 irradiation; and more recently in the analysis, evaluation and interpretation of the data, which was supported by the DOE Office of Fusion Energy Sciences. Finally, we thank Dr. Jia Hong Ke for his assistance in including the modeling comparison.